\documentclass[aps,prl,twocolumn,superscriptaddress]{revtex4-1}

\usepackage{graphicx}
\usepackage{bm}
\usepackage{amsmath}
\usepackage{siunitx}
\usepackage{amssymb}
\begin{document}



\title{Low Photon Count Phase Retrieval Using Deep Learning}



\author{Alexandre Goy}
\email{agoy@mit.edu}
\author{Kwabena Arthur}
\author{Shuai Li}
\author{George Barbastathis}
\altaffiliation{Singapore-MIT Alliance for Research and Technology (SMART) Centre, Singapore 117543, Singapore.}
\affiliation{Mechanical Engineering, Massachusetts Institute of Technology, Cambridge, MA02139, USA.}



\date{\today}

\begin{abstract}
Imaging systems' performance at low light intensity is affected by shot noise, which becomes increasingly strong as the power of the light source decreases. In this paper we experimentally demonstrate the use of deep neural networks to recover objects illuminated with weak light and demonstrate better performance than with the classical Gerchberg-Saxton phase retrieval algorithm for equivalent signal over noise ratio. Prior knowledge about the object is implicitly contained in the training data set and feature detection is possible for a signal over noise ratio close to one. We apply this principle to a phase retrieval problem and show successful recovery of the object's most salient features with as little as one photon per detector pixel on average in the illumination beam. We also show that the phase reconstruction is significantly improved by training the neural network with an initial estimate of the object, as opposed as training it with the raw intensity measurement.
\end{abstract}


\maketitle


Many imaging systems only yield partial or distorted information about the object being imaged. Typical causes include loss of spatial frequencies, lack of phase information, unknown scatterers in the optical train, aberrations, and noise in the illumination or detection. In these situations, the mathematical operator describing the imaging system becomes ill-posed and usually requires regularization. A regularizer is an operator designed to favor solutions that match our prior knowledge about the object, if any. The choice of the regularizer itself is often arbitrary and based on practical experience. Recently, Deep Neural Networks (DNNs) have attracted much attention in the field of computational imaging, for they  provide a way to regularize a problem adaptively. As of today, DNNs have been proven efficient solvers in many imaging applications such as deblurring~\cite{xu:2014}, undersampled imaging~\cite{mardani:2017}, ghost imaging~\cite{lyu:2017b}, phase retrieval~\cite{sinha:2017, li:2017, rivenson:2018, metzler:2018, kemp:2018, boominathan:2018}, adaptive illumination microscopy~\cite{horstmayer:2017}, adaptive optics~\cite{sandler:1991}, and optical tomography~\cite{kamilov:2015, kamilov:2016}. In particular, in the context of phase retrieval, it has been demonstrated numerically that machine learning can improve the imaging condition~\cite{metzler:2018}.

In this paper, we demonstrate experimentally for the first time, to our knowledge, that DNNs can solve a coherent phase retrieval problem affected by strong shot noise at various levels. We also provide corresponding numerical simulations. In situations where the light source is weak, the detection signal to noise ratio (SNR) is ultimately limited by the quantized nature of light. Because of its fundamental nature, shot noise cannot be avoided and regularization schemes must be devised to handle it. As the noise becomes more significant, reconstruction algorithms' performance in general deteriorates; this is the regime where we expect the biggest payoff from the DNN, assuming that it has been successfully trained to recover the object features that best explain the observed signal distribution. Best results are obtained for objects within restricted classes, {\it i.e.} sharing similar constrained features, or equivalently having a sparse description in some domain of appropriately chosen basis functions. To illustrate this, we used two sets of databases to train DNNs: a relatively restricted class of Integrated Circuit (IC) layouts, and the more general ImageNet~\cite{imagenet:2009} image dataset. We found that the DNN reconstructions attain better visual quality for IC layouts at low photon counts (one$\sim$two per pixel per frame) than for ImageNet.

DNNs represent a very versatile method for inferring the relationship between objects and their corresponding measurements through the imaging system. A DNN is typically trained on a set of examples, each example containing the ideal image of the object (the ground truth) and a corresponding measurement. The DNN can be viewed as an operator mapping the measurement (or a known function of the measurement) to the desired image. The internal parameters of the DNN are adjusted to minimize a loss function that describes how close the image is to the ground truth. After the training, examples from a test set, which have not been used in the training phase, are given to the DNN, which then outputs the reconstructed images.

The phase retrieval problem addressed in this work can be written, for an optically thin object, as:
\begin{equation}\label{equ:phase_retrieval}
\mathbf{g}(x, y) = \left|F_{L}\left[u_{\mathrm{inc}}(x, y)\mathbf{t}(x, y)e^{j\mathbf{f}(x, y)}\right]\right|^2,
\end{equation}
\noindent where $(x, y)$ are the lateral coordinates, $\mathbf{g}$ is the intensity measurement in the detector plane, $\mathbf{t}$ and $\mathbf{f}$ are, respectively, the modulus and phase of the field immediately after the object, $u_{\mathrm{inc}}$ the incident field in the object plane, and $F_L$ the Fresnel propagation operator over a distance $L$. In what follows, we assume that the object modulates only the phase, therefore $\mathbf{t}(x, y) = 1$, and we define: $\mathbf{g} = H(\mathbf{f})$. The optimization problem implicitly solved by the DNN can be written as: 
\begin{equation}\label{equ:optimization}
\mathbf{\hat{f}} = \operatorname*{argmin}_\mathbf{f} \:\: \psi\left\{\rule[-1ex]{0cm}{3ex} H(\mathbf{f}), \mathbf{g}, \bm{\Theta}(\mathbf{f})\right\},
\end{equation}
\noindent where $\psi$ is the functional to minimize and $\bm{\Theta}$ the regularizer operating on $\mathbf{f}$, {\it i.e.} imposing constraints on the solution. In a classical optimization procedure, the regularizer would be chosen \textit{ad hoc}. Instead, here we let the DNN discover a regularization adapted to the specific class of objects we train with. 

In this work, the loss function is chosen as the negative Pearson correlation coefficient (NPCC) defined in the appendix. The use of the NPCC as a loss function, as opposed for example to the mean square error, proved to be a better metric for DNN training in the context of phase retrieval, especially with sparse objects~\cite{li:2017}.

For our phase retrieval problem, one possibility is to train the DNN with $\left(\mathbf{f}_k, \mathbf{g}_k\right)$ couples, $k$ being the index within the training set. We refer to this approach as the ``end-to-end'' method as it makes use of the endpoints of the optical system, {\it i.e.} the object phase $\mathbf{f}$ and raw intensity measurement $\mathbf{g}$. It should be noted that, in the end-to-end method, in addition to the regularization, the DNN carries the burden of learning the law of Fresnel propagation. Since Fresnel propagation is a well characterized physical law, it seems inefficient to have the DNN being optimized, even partially, to explain it. Some knowledge about the physical laws has to be included in the training process in order for the DNN to focus on learning a regularizer. 

The phase retrieval problem described in Eq.~\ref{equ:phase_retrieval} cannot be inverted directly, simply because the detector is not sensitive to phase. Therefore, there is no unique way of disentangling the contribution of the physics and the contribution of the noise (or any other stochastic process involved). However, the well-known Gerchberg-Saxton (GS)~\cite{gerchberg:1971} and the gradient descent algorithms for phase retrieval provide a useful insight. Even though the phase is not known in the detector plane, an approximate phase can be assumed and used to project the field back to the object plane using the inverse Fresnel operator. In this work, we associate the phase of the incident beam in the detector plane with the square root of the intensity measurement to produce a complex field, which is propagated back to the object plane. The phase of this complex field in the object plane is referred to as an ``approximant'' (or GS-approximant as it is inspired by the GS algorithm) as it is generally closer to the solution than the raw intensity measurement. Note that the adjoint of operator $H$, used in the gradient descent method, can also be used to generate an approximant, however, we will restrict our analysis to the GS-approximant. The approximant can be used \textit{in lieu} of the raw measurement for the DNN training. This is an example of a ``physics-informed'' method as part of the physical process is embedded in the approximant itself. A similar procedure involving such a preprocessing step has been described recently in~\cite{sun:2018}.

\begin{figure}
\includegraphics{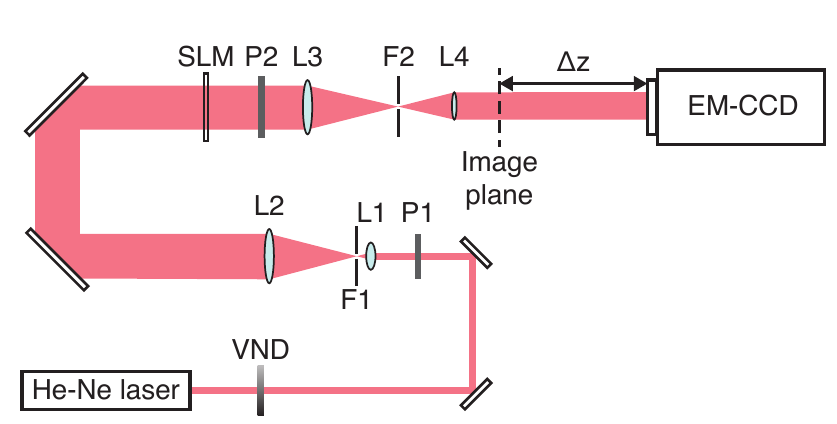}
\caption{\label{fig:opt_setup} Optical apparatus. VND: variable neutral density filter, P1-P2: polarizers, L1: 10$\times$ objective, L2: $\SI{100}{\milli\meter}$ lens, L3: $\SI{230}{\milli\meter}$ lens, L4: $\SI{100}{\milli\meter}$ lens, F1: $\SI{5}{\micro\meter}$ pinhole, F2: iris. SLM: transmissive spatial light modulator.}
\end{figure}

In what follows, we describe a series of experiments designed to systematically compare the end-to-end, physics-informed (using the GS-approximant), and the classical Gerchberg-Saxton methods for different levels of noise. Corresponding simulations have been performed and are presented in the supplementary material. The experimental apparatus is depicted in Fig.~\ref{fig:opt_setup}.

The light source is a Helium-Neon laser emitting continuous wave radiation at $\SI{632.8}{\nano\meter}$. The laser beam intensity is controlled by a calibrated variable neutral density filter. The beam is then spatially filtered, expanded and passed through a transmissive spatial light modulator (SLM) (Holoeye LC2012) with $\SI{36}{\micro\meter}$ square pixels. In order to maximize the SLM phase modulation capability, the incident light is linearly polarized (P1) at a certain angle (45$^\circ$ from the horizontal axis). The modulated light from the SLM is filtered by a second polarizer (P2). The complex (phase and intensity) transmittance of the SLM was calibrated interferometrically for the particular polarizers configuration used in the experiment. The SLM surface is reduced by a factor of 2.3 by a telescope system (lenses L3 and L4 in Fig.~\ref{fig:opt_setup}) in order for the diffracted pattern to fit within the detector. The detector is an EM-CCD 1004$\times$1002 array (QImaging Rolera EM-C2) of $8\times\SI{8}{\micro\meter}$ pixels. The EM gain and exposure time of the camera are controlled by software. The detector is placed at a distance $\Delta z = \SI{400}{\milli\meter}$ from the image plane. An additional neutral density filter with an optical density of 2 is placed in front of the detector to suppress background light and adjust the photon level range. The actual optical power is measured between filter F2 and lens L4 with a Silicon detector. Details about the calibraton are given in the supplementary material. It should be noted that the SLM has a residual intensity modulation effect, which was measured during the calibration step (see supplementary material). The DNN is trained to recover the phase component only. The Gerchberg-Saxton algorithm is run by assuming the intensity in the SLM plane as being that of the incident beam. The known relationship between amplitude and phase could have been used to constrain the convergence of the GS algorithm, but such constraint was not made available to the DNN either.

\begin{figure*}
\includegraphics{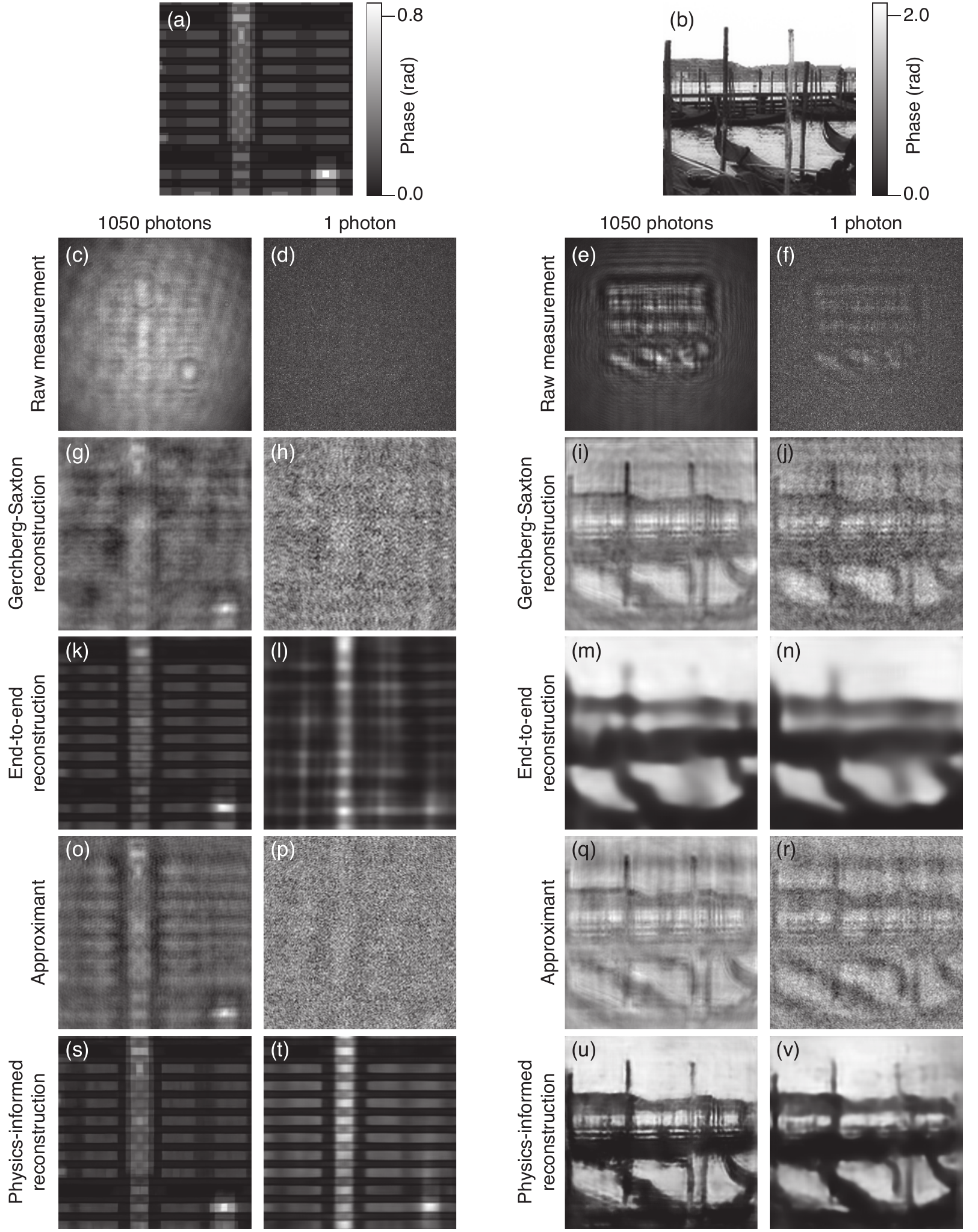}
\caption{\label{fig:images} \mbox{(a-b)} Ground truth phase of one example from each test set of IC layouts and ImageNet. \mbox{(c-f)} Raw measurements in the detector plane. \mbox{(g-j)} Gerchberg-Saxton algorithm reconstructions from the raw measurements c-f. \mbox{(k-n)} DNN reconstructions with the end-to-end method. \mbox{(o-r)} Approximants in the image plane. \mbox{(s-v)} DNN reconstructions from the approximants o-r with the physics-informed method. For better display, the grayscales of all images have been normalized to range from the minimal to the maximal value. Images a, b and g to v represent a phase in the image plane and have a physical size of 4$\times$4mm, while images c to f represent an intensity in the detector plane and have a physical size of 8$\times$8mm.}
\end{figure*}

For each image category (ImageNet and IC layouts) and for each noise level, a different DNN is trained. The examples are split into a training set, a validation set and a test set containing 9,500, 450 and 50 examples of 256 by 256 pixel images, respectively. The DNN input and output images are 256 by 256 pixel, which is the native resolution of the images in the dataset. The ground truth images displayed on the SLM are also 256 by 256 pixels. The detector images are 1004 by 1002 pixels, therefore they must be resampled. For the end-to-end method, the detector images are interpolated from 1002 by 1002 to 256 by 256 using bilinear interpolation. For the physics-informed method, each detector image is zero-padded to a size such that the inverse Fresnel propagator would yield an approximant in which the object covered a 256 by 256 pixel area. This procedure has the advantage of performing the Fresnel propagation and the resampling of the image in a single operation. The DNN has the same encoder-decoder architecture as presented in~\cite{li:2017} except that five instead of six convolutional layers are used in the encoder and decoder parts.

\begin{figure}
\includegraphics{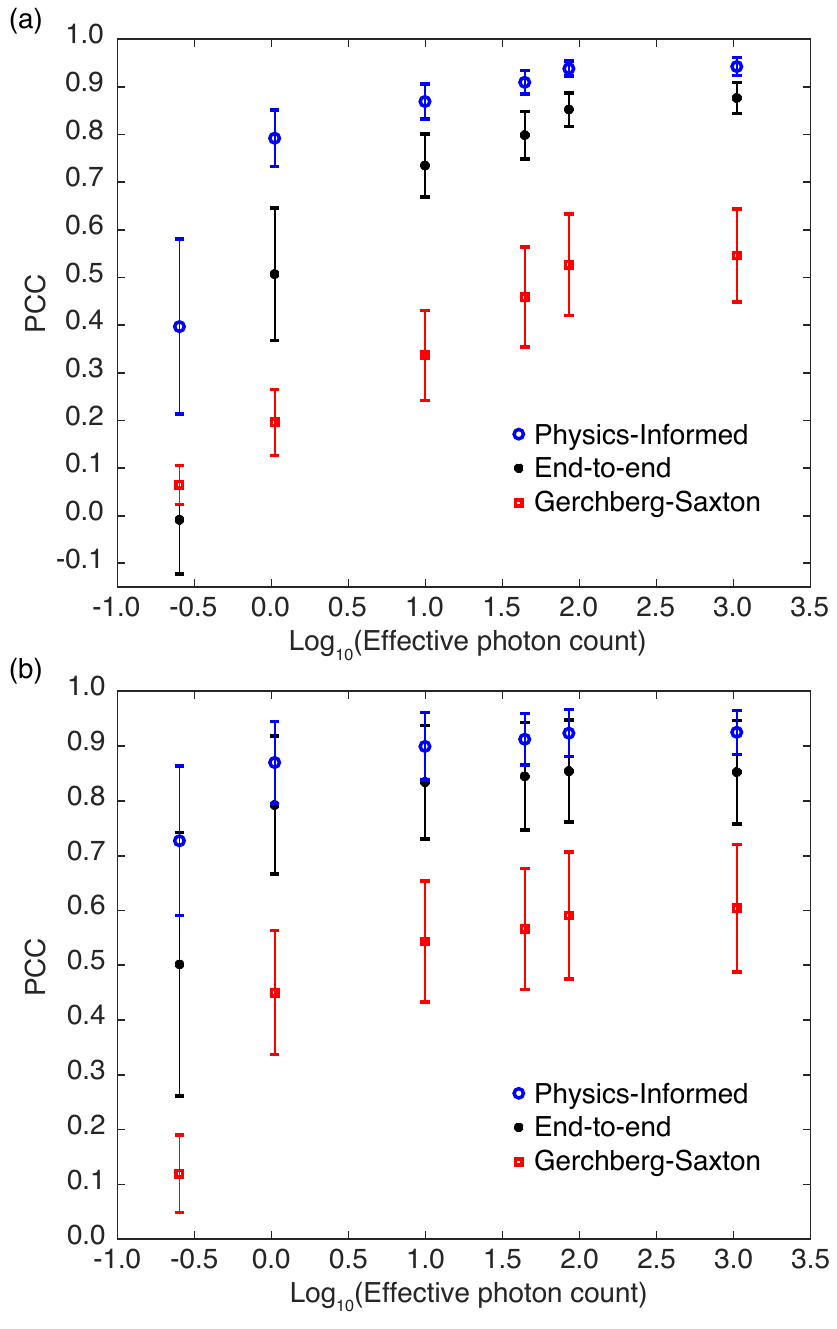}
\caption{\label{fig:correlations} Pearson correlation coefficient between the ground truth and the DNN reconstructions. (a) IC layout data set. (b) ImageNet data set. The markers indicate the mean over the test set (50 examples) and the error bars $\pm 1$ standard deviation from the mean.}
\end{figure}

Examples of reconstruction from the test sets for both ImageNet and IC layouts are shown in Fig.~\ref{fig:images} for two extreme photon level cases. Table~\ref{tab:noiselevels} summarizes the noise level for each expriment shown in Fig.~\ref{fig:images} and~\ref{fig:correlations}. The noise levels indicated in the table refer to the incident beam, {\it i.e.} with no modulation on the SLM. When a pattern is displayed on the SLM, the SNR at the detector plane varies strongly spatially as a result of intensity redistribution, which is why using the incident beam as reference was preferred. The integration time was set at $\SI{2}{\milli\second}$ for all experiments mentiond in Table~\ref{tab:noiselevels} and Fig.~\ref{fig:images} and~\ref{fig:correlations}. The integration time was kept short to avoid degradation of the SNR due to air turbulence.

\begin{table}[b]
\caption{\label{tab:noiselevels} Noise levels and photon count for the experiments shown in Fig.~\ref{fig:correlations}. The illumination conditions are the same for both the IC layout and the ImageNet datasets. The photon count is the effective number of photons after dividing by the quantum efficienty per detector pixel averaged over the whole detector field for the incident beam (no modulation on the SLM). The procedure for measuring the photon count is given in the supplementary material. The SNR is the mean of the incident beam signal divided by its standard deviation and averaged over the whole field of view. The limit SNR is the square root of the number of photons.}
\begin{tabular}{|c|c|c|c|c|} \hline
Experiment & EM gain & Photon count $\pm 5\%$ & SNR & Limit SNR \\ \hline\hline
1 & 1 & 1050 & 20 & 32 \\ \hline
2 & 1 & 85 & 2.7 & 9.2 \\ \hline
3 & 1 & 44 & 1.45 & 6.6 \\ \hline
4 & 4.8 & 9.9 & 0.9 & 3.1 \\ \hline
5 & 54 & 1.1 & 0.5 & 1.0 \\ \hline
6 & 54 & 0.25 & 0.24 & 0.5 \\ \hline
\end{tabular}
\end{table}

The results shown in Fig.~\ref{fig:images} allow us to draw qualitative conclusions. As can be seen in Fig.~\ref{fig:images} (g-j) and (o-r), the DNN is very efficient in suppressing the granularity typical of shot noise. The end-to-end method reconstructions appear as low-pass filtered versions of the original image, especially for ImageNet examples. IC layout examples are still reconstructed with sharp edges as this feature is omnipresent in the IC layout. The interpretation is that the DNN does not fully learn the diffraction operator, but rather learns how to suppress fringes and other diffraction related patterns and also how to promote characteristic features of the training examples. The physics-informed reconstructions are visually better because, in this case, high frequencies are provided to the DNN by the approximant (especially visible in Fig.~\ref{fig:images}q). In the low photon example of the IC layout (Fig.~\ref{fig:images}t), the general pattern is recovered, but additional spurious tracks have been added by the DNN that seems to promote periodicity, a feature quite prominent in IC layout examples.

We use the Pearson correlation coefficient ($\text{PCC} = -\text{NPCC}$) as a figure of merit for the quality of the reconstructions; the results are shown in Fig.~\ref{fig:correlations}. In the case of the IC layout, for all photon levels, the physics-informed method performs systematically better than the end-to-end method, which in turn performs better than the GS algorithm. A similar result holds for the ImageNet example set, except that there is less difference between the end-to-end and the physics-informed reconstruction and also that the standard deviation of the reconstruction quality is larger even for high photon levels. The GS reconstructions for high photon level do not display this trend (their standard deviation remains equally large). This latter observation confirms that the strong prior in the IC layout geometry is efficiently exploited by the DNN.

The PCC is not sensitive to the magnitude of the images ({\it i.e.} PCC$(A, B)$ = PCC$(\alpha A, \beta B), \alpha, \beta \in \mathbb{R}$), the phase images are thus reconstructed up to a scaling factor. However, for a given DNN the scaling factor is constant and can be retrieved by comparing the validation set ground truth examples and corresponding reconstructions. In practice, the scaling factor is obtained by comapring the histograms of the ground truths and reconstructions images.


The approximant clearly helps in recovering high fidelity images. The question of knowing what is the best way of obtaining an approximant in the context of phase retrieval is beyond the scope of this paper. It should be recognized that the GS-approximant the way it is computed here corresponds to half of the first iteration of the GS algorithm. The question whether it is worthy to iterate more in order to generate an approximant is still open, but preliminary results tend to show that little is gained by iterating more.

This work was supported by the Intelligence Advanced Research Projects Activity (IARPA) FA8650-17-C-9113.

\appendix*

\section{Negative Pearson correlation coefficient}

For two images $A$ and $B$, with pixels indexed by $i$ and, the negative Pearson correlation coefficient (NPCC) is defined as:
\begin{equation}\label{equ:npcc}
\mathrm{NPCC}(A, B) = \frac{-\sum_{i}(A_{i}-\bar{A})(B_{i}-\bar{B})}{\sqrt{\sum_{i}(A_{i}-\bar{A})^2\sum_{i}(B_{i}-\bar{B})^2}},\\
\end{equation}
\noindent where the bar denotes the average. The NPCC reaches -1 for a perfect match and higher values otherwise.

\bibliography{2018-05-25LowPhotonCount}

\providecommand{\noopsort}[1]{}\providecommand{\singleletter}[1]{#1}%
\begin{thebibliography}{16}%
\makeatletter
\providecommand \@ifxundefined [1]{%
 \@ifx{#1\undefined}
}%
\providecommand \@ifnum [1]{%
 \ifnum #1\expandafter \@firstoftwo
 \else \expandafter \@secondoftwo
 \fi
}%
\providecommand \@ifx [1]{%
 \ifx #1\expandafter \@firstoftwo
 \else \expandafter \@secondoftwo
 \fi
}%
\providecommand \natexlab [1]{#1}%
\providecommand \enquote  [1]{``#1''}%
\providecommand \bibnamefont  [1]{#1}%
\providecommand \bibfnamefont [1]{#1}%
\providecommand \citenamefont [1]{#1}%
\providecommand \href@noop [0]{\@secondoftwo}%
\providecommand \href [0]{\begingroup \@sanitize@url \@href}%
\providecommand \@href[1]{\@@startlink{#1}\@@href}%
\providecommand \@@href[1]{\endgroup#1\@@endlink}%
\providecommand \@sanitize@url [0]{\catcode `\\12\catcode `\$12\catcode
  `\&12\catcode `\#12\catcode `\^12\catcode `\_12\catcode `\%12\relax}%
\providecommand \@@startlink[1]{}%
\providecommand \@@endlink[0]{}%
\providecommand \url  [0]{\begingroup\@sanitize@url \@url }%
\providecommand \@url [1]{\endgroup\@href {#1}{\urlprefix }}%
\providecommand \urlprefix  [0]{URL }%
\providecommand \Eprint [0]{\href }%
\providecommand \doibase [0]{http://dx.doi.org/}%
\providecommand \selectlanguage [0]{\@gobble}%
\providecommand \bibinfo  [0]{\@secondoftwo}%
\providecommand \bibfield  [0]{\@secondoftwo}%
\providecommand \translation [1]{[#1]}%
\providecommand \BibitemOpen [0]{}%
\providecommand \bibitemStop [0]{}%
\providecommand \bibitemNoStop [0]{.\EOS\space}%
\providecommand \EOS [0]{\spacefactor3000\relax}%
\providecommand \BibitemShut  [1]{\csname bibitem#1\endcsname}%
\let\auto@bib@innerbib\@empty
\bibitem [{\citenamefont {Xu}\ \emph {et~al.}(2014)\citenamefont {Xu},
  \citenamefont {Ren}, \citenamefont {Liu},\ and\ \citenamefont
  {Jia}}]{xu:2014}%
  \BibitemOpen
  \bibfield  {author} {\bibinfo {author} {\bibfnamefont {L.}~\bibnamefont
  {Xu}}, \bibinfo {author} {\bibfnamefont {J.~S.}\ \bibnamefont {Ren}},
  \bibinfo {author} {\bibfnamefont {C.}~\bibnamefont {Liu}}, \ and\ \bibinfo
  {author} {\bibfnamefont {J.}~\bibnamefont {Jia}},\ }in\ \href
  {http://papers.nips.cc/paper/5485-deep-convolutional-neural-network-for-image-deconvolution.pdf}
  {\emph {\bibinfo {booktitle} {Advances in Neural Information Processing
  Systems 27}}},\ \bibinfo {editor} {edited by\ \bibinfo {editor}
  {\bibfnamefont {Z.}~\bibnamefont {Ghahramani}}, \bibinfo {editor}
  {\bibfnamefont {M.}~\bibnamefont {Welling}}, \bibinfo {editor} {\bibfnamefont
  {C.}~\bibnamefont {Cortes}}, \bibinfo {editor} {\bibfnamefont {N.~D.}\
  \bibnamefont {Lawrence}}, \ and\ \bibinfo {editor} {\bibfnamefont {K.~Q.}\
  \bibnamefont {Weinberger}}}\ (\bibinfo  {publisher} {Curran Associates,
  Inc.},\ \bibinfo {year} {2014})\ pp.\ \bibinfo {pages}
  {1790--1798}\BibitemShut {NoStop}%
\bibitem [{\citenamefont {Mardani}\ \emph {et~al.}(2017)\citenamefont
  {Mardani}, \citenamefont {Monajemi}, \citenamefont {Papyan}, \citenamefont
  {Vasanawala}, \citenamefont {Donoho},\ and\ \citenamefont
  {Pauly}}]{mardani:2017}%
  \BibitemOpen
  \bibfield  {author} {\bibinfo {author} {\bibfnamefont {M.}~\bibnamefont
  {Mardani}}, \bibinfo {author} {\bibfnamefont {H.}~\bibnamefont {Monajemi}},
  \bibinfo {author} {\bibfnamefont {V.}~\bibnamefont {Papyan}}, \bibinfo
  {author} {\bibfnamefont {S.}~\bibnamefont {Vasanawala}}, \bibinfo {author}
  {\bibfnamefont {D.}~\bibnamefont {Donoho}}, \ and\ \bibinfo {author}
  {\bibfnamefont {J.}~\bibnamefont {Pauly}},\ }\href@noop {} {\bibfield
  {journal} {\bibinfo  {journal} {Archive}\ ,\ \bibinfo {pages}
  {arXiv:1711.10046v1}} (\bibinfo {year} {2017})}\BibitemShut {NoStop}%
\bibitem [{\citenamefont {Lyu}\ \emph {et~al.}(2017)\citenamefont {Lyu},
  \citenamefont {Wang}, \citenamefont {Wang}, \citenamefont {Wang},
  \citenamefont {Li}, \citenamefont {Chen},\ and\ \citenamefont
  {Situ}}]{lyu:2017b}%
  \BibitemOpen
  \bibfield  {author} {\bibinfo {author} {\bibfnamefont {M.}~\bibnamefont
  {Lyu}}, \bibinfo {author} {\bibfnamefont {W.}~\bibnamefont {Wang}}, \bibinfo
  {author} {\bibfnamefont {H.}~\bibnamefont {Wang}}, \bibinfo {author}
  {\bibfnamefont {H.}~\bibnamefont {Wang}}, \bibinfo {author} {\bibfnamefont
  {G.}~\bibnamefont {Li}}, \bibinfo {author} {\bibfnamefont {N.}~\bibnamefont
  {Chen}}, \ and\ \bibinfo {author} {\bibfnamefont {G.}~\bibnamefont {Situ}},\
  }\href@noop {} {\bibfield  {journal} {\bibinfo  {journal} {Scientific
  Reportsvolume}\ }\textbf {\bibinfo {volume} {7}},\ \bibinfo {pages} {17865}
  (\bibinfo {year} {2017})}\BibitemShut {NoStop}%
\bibitem [{\citenamefont {Sinha}\ \emph {et~al.}(2017)\citenamefont {Sinha},
  \citenamefont {Lee}, \citenamefont {Li}, ,\ and\ \citenamefont
  {Barbastathis}}]{sinha:2017}%
  \BibitemOpen
  \bibfield  {author} {\bibinfo {author} {\bibfnamefont {A.}~\bibnamefont
  {Sinha}}, \bibinfo {author} {\bibfnamefont {J.}~\bibnamefont {Lee}}, \bibinfo
  {author} {\bibfnamefont {S.}~\bibnamefont {Li}}, , \ and\ \bibinfo {author}
  {\bibfnamefont {G.}~\bibnamefont {Barbastathis}},\ }\href@noop {} {\bibfield
  {journal} {\bibinfo  {journal} {Optica}\ }\textbf {\bibinfo {volume} {4}},\
  \bibinfo {pages} {1117} (\bibinfo {year} {2017})}\BibitemShut {NoStop}%
\bibitem [{\citenamefont {Li}\ \emph {et~al.}(2017)\citenamefont {Li},
  \citenamefont {Deng}, \citenamefont {Lee}, \citenamefont {Sinha},\ and\
  \citenamefont {Barbastathis}}]{li:2017}%
  \BibitemOpen
  \bibfield  {author} {\bibinfo {author} {\bibfnamefont {S.}~\bibnamefont
  {Li}}, \bibinfo {author} {\bibfnamefont {M.}~\bibnamefont {Deng}}, \bibinfo
  {author} {\bibfnamefont {J.}~\bibnamefont {Lee}}, \bibinfo {author}
  {\bibfnamefont {A.}~\bibnamefont {Sinha}}, \ and\ \bibinfo {author}
  {\bibfnamefont {G.}~\bibnamefont {Barbastathis}},\ }\href@noop {} {\bibfield
  {journal} {\bibinfo  {journal} {Archive}\ ,\ \bibinfo {pages}
  {arXiv:1711.06810}} (\bibinfo {year} {2017})}\BibitemShut {NoStop}%
\bibitem [{\citenamefont {Rivenson}\ \emph {et~al.}(2018)\citenamefont
  {Rivenson}, \citenamefont {Zhang}, \citenamefont {Günaydın}, \citenamefont
  {Teng},\ and\ \citenamefont {Ozcan}}]{rivenson:2018}%
  \BibitemOpen
  \bibfield  {author} {\bibinfo {author} {\bibfnamefont {Y.}~\bibnamefont
  {Rivenson}}, \bibinfo {author} {\bibfnamefont {Y.}~\bibnamefont {Zhang}},
  \bibinfo {author} {\bibfnamefont {H.}~\bibnamefont {Günaydın}}, \bibinfo
  {author} {\bibfnamefont {D.}~\bibnamefont {Teng}}, \ and\ \bibinfo {author}
  {\bibfnamefont {A.}~\bibnamefont {Ozcan}},\ }\href@noop {} {\bibfield
  {journal} {\bibinfo  {journal} {Light: Science and Applications}\ }\textbf
  {\bibinfo {volume} {7}},\ \bibinfo {pages} {17141} (\bibinfo {year}
  {2018})}\BibitemShut {NoStop}%
\bibitem [{\citenamefont {Metzler}\ \emph {et~al.}(2018)\citenamefont
  {Metzler}, \citenamefont {Schniter}, \citenamefont {Veeraraghavan},\ and\
  \citenamefont {Baraniuk}}]{metzler:2018}%
  \BibitemOpen
  \bibfield  {author} {\bibinfo {author} {\bibfnamefont {C.~A.}\ \bibnamefont
  {Metzler}}, \bibinfo {author} {\bibfnamefont {P.}~\bibnamefont {Schniter}},
  \bibinfo {author} {\bibfnamefont {A.}~\bibnamefont {Veeraraghavan}}, \ and\
  \bibinfo {author} {\bibfnamefont {R.~G.}\ \bibnamefont {Baraniuk}},\
  }\href@noop {} {\bibfield  {journal} {\bibinfo  {journal} {Archive}\ ,\
  \bibinfo {pages} {arXiv:1803.00212}} (\bibinfo {year} {2018})}\BibitemShut
  {NoStop}%
\bibitem [{\citenamefont {Kemp}(2018)}]{kemp:2018}%
  \BibitemOpen
  \bibfield  {author} {\bibinfo {author} {\bibfnamefont {Z.~D.~C.}\
  \bibnamefont {Kemp}},\ }\href@noop {} {\bibfield  {journal} {\bibinfo
  {journal} {Archive}\ ,\ \bibinfo {pages} {arXiv:1709.09940}} (\bibinfo {year}
  {2018})}\BibitemShut {NoStop}%
\bibitem [{\citenamefont {Boominathan}\ \emph {et~al.}(2018)\citenamefont
  {Boominathan}, \citenamefont {Maniparambil}, \citenamefont {Gupta},
  \citenamefont {Baburajan},\ and\ \citenamefont {Mitra}}]{boominathan:2018}%
  \BibitemOpen
  \bibfield  {author} {\bibinfo {author} {\bibfnamefont {L.}~\bibnamefont
  {Boominathan}}, \bibinfo {author} {\bibfnamefont {M.}~\bibnamefont
  {Maniparambil}}, \bibinfo {author} {\bibfnamefont {H.}~\bibnamefont {Gupta}},
  \bibinfo {author} {\bibfnamefont {R.}~\bibnamefont {Baburajan}}, \ and\
  \bibinfo {author} {\bibfnamefont {K.}~\bibnamefont {Mitra}},\ }\href@noop {}
  {\bibfield  {journal} {\bibinfo  {journal} {Archive}\ ,\ \bibinfo {pages}
  {arXiv:1805.03593}} (\bibinfo {year} {2018})}\BibitemShut {NoStop}%
\bibitem [{\citenamefont {Horstmeyer}\ \emph {et~al.}(2017)\citenamefont
  {Horstmeyer}, \citenamefont {Chen}, \citenamefont {Kappes},\ and\
  \citenamefont {Judkewitz}}]{horstmayer:2017}%
  \BibitemOpen
  \bibfield  {author} {\bibinfo {author} {\bibfnamefont {R.}~\bibnamefont
  {Horstmeyer}}, \bibinfo {author} {\bibfnamefont {R.~Y.}\ \bibnamefont
  {Chen}}, \bibinfo {author} {\bibfnamefont {B.}~\bibnamefont {Kappes}}, \ and\
  \bibinfo {author} {\bibfnamefont {B.}~\bibnamefont {Judkewitz}},\ }\href@noop
  {} {\bibfield  {journal} {\bibinfo  {journal} {Archive}\ ,\ \bibinfo {pages}
  {arXiv:1709.07223v1}} (\bibinfo {year} {2017})}\BibitemShut {NoStop}%
\bibitem [{\citenamefont {Sandler}\ \emph {et~al.}(1991)\citenamefont
  {Sandler}, \citenamefont {Barrett}, \citenamefont {Palmer}, \citenamefont
  {Fugate},\ and\ \citenamefont {Wild}}]{sandler:1991}%
  \BibitemOpen
  \bibfield  {author} {\bibinfo {author} {\bibfnamefont {D.~G.}\ \bibnamefont
  {Sandler}}, \bibinfo {author} {\bibfnamefont {T.~K.}\ \bibnamefont
  {Barrett}}, \bibinfo {author} {\bibfnamefont {D.~A.}\ \bibnamefont {Palmer}},
  \bibinfo {author} {\bibfnamefont {R.~Q.}\ \bibnamefont {Fugate}}, \ and\
  \bibinfo {author} {\bibfnamefont {W.~J.}\ \bibnamefont {Wild}},\ }\href@noop
  {} {\bibfield  {journal} {\bibinfo  {journal} {Nature}\ }\textbf {\bibinfo
  {volume} {351}},\ \bibinfo {pages} {300} (\bibinfo {year}
  {1991})}\BibitemShut {NoStop}%
\bibitem [{\citenamefont {Kamilov}\ \emph {et~al.}(2015)\citenamefont
  {Kamilov}, \citenamefont {Papadopoulos}, \citenamefont {Shoreh},
  \citenamefont {Goy}, \citenamefont {Vonesch}, \citenamefont {Unser},\ and\
  \citenamefont {Psaltis}}]{kamilov:2015}%
  \BibitemOpen
  \bibfield  {author} {\bibinfo {author} {\bibfnamefont {U.~S.}\ \bibnamefont
  {Kamilov}}, \bibinfo {author} {\bibfnamefont {I.~N.}\ \bibnamefont
  {Papadopoulos}}, \bibinfo {author} {\bibfnamefont {M.~H.}\ \bibnamefont
  {Shoreh}}, \bibinfo {author} {\bibfnamefont {A.}~\bibnamefont {Goy}},
  \bibinfo {author} {\bibfnamefont {C.}~\bibnamefont {Vonesch}}, \bibinfo
  {author} {\bibfnamefont {M.}~\bibnamefont {Unser}}, \ and\ \bibinfo {author}
  {\bibfnamefont {D.}~\bibnamefont {Psaltis}},\ }\href@noop {} {\bibfield
  {journal} {\bibinfo  {journal} {Optica}\ }\textbf {\bibinfo {volume} {2}},\
  \bibinfo {pages} {517} (\bibinfo {year} {2015})}\BibitemShut {NoStop}%
\bibitem [{\citenamefont {Kamilov}\ \emph {et~al.}(2016)\citenamefont
  {Kamilov}, \citenamefont {Papadopoulos}, \citenamefont {Shoreh},
  \citenamefont {Goy}, \citenamefont {Vonesch}, \citenamefont {Unser},\ and\
  \citenamefont {Psaltis}}]{kamilov:2016}%
  \BibitemOpen
  \bibfield  {author} {\bibinfo {author} {\bibfnamefont {U.~S.}\ \bibnamefont
  {Kamilov}}, \bibinfo {author} {\bibfnamefont {I.~N.}\ \bibnamefont
  {Papadopoulos}}, \bibinfo {author} {\bibfnamefont {M.~H.}\ \bibnamefont
  {Shoreh}}, \bibinfo {author} {\bibfnamefont {A.}~\bibnamefont {Goy}},
  \bibinfo {author} {\bibfnamefont {C.}~\bibnamefont {Vonesch}}, \bibinfo
  {author} {\bibfnamefont {M.}~\bibnamefont {Unser}}, \ and\ \bibinfo {author}
  {\bibfnamefont {D.}~\bibnamefont {Psaltis}},\ }\href@noop {} {\bibfield
  {journal} {\bibinfo  {journal} {IEEE Transactions on Computational Imaging}\
  }\textbf {\bibinfo {volume} {2}},\ \bibinfo {pages} {59} (\bibinfo {year}
  {2016})}\BibitemShut {NoStop}%
\bibitem [{\citenamefont {Deng}\ \emph {et~al.}(2009)\citenamefont {Deng},
  \citenamefont {Dong}, \citenamefont {Socher}, \citenamefont {Li},
  \citenamefont {Li},\ and\ \citenamefont {Fei-Fei}}]{imagenet:2009}%
  \BibitemOpen
  \bibfield  {author} {\bibinfo {author} {\bibfnamefont {J.}~\bibnamefont
  {Deng}}, \bibinfo {author} {\bibfnamefont {W.}~\bibnamefont {Dong}}, \bibinfo
  {author} {\bibfnamefont {R.}~\bibnamefont {Socher}}, \bibinfo {author}
  {\bibfnamefont {L.~J.}\ \bibnamefont {Li}}, \bibinfo {author} {\bibfnamefont
  {K.}~\bibnamefont {Li}}, \ and\ \bibinfo {author} {\bibfnamefont
  {L.}~\bibnamefont {Fei-Fei}},\ }in\ \href {\doibase
  10.1109/CVPR.2009.5206848} {\emph {\bibinfo {booktitle} {2009 IEEE Conference
  on Computer Vision and Pattern Recognition}}}\ (\bibinfo {year} {2009})\ pp.\
  \bibinfo {pages} {248--255}\BibitemShut {NoStop}%
\bibitem [{\citenamefont {Gerchberg}\ and\ \citenamefont
  {Saxton}(1971)}]{gerchberg:1971}%
  \BibitemOpen
  \bibfield  {author} {\bibinfo {author} {\bibfnamefont {R.~W.}\ \bibnamefont
  {Gerchberg}}\ and\ \bibinfo {author} {\bibfnamefont {W.~O.}\ \bibnamefont
  {Saxton}},\ }\href@noop {} {\bibfield  {journal} {\bibinfo  {journal}
  {Optik}\ }\textbf {\bibinfo {volume} {35}},\ \bibinfo {pages} {237} (\bibinfo
  {year} {1971})}\BibitemShut {NoStop}%
\bibitem [{\citenamefont {Sun}\ \emph {et~al.}(2018)\citenamefont {Sun},
  \citenamefont {Xia},\ and\ \citenamefont {Kamilov}}]{sun:2018}%
  \BibitemOpen
  \bibfield  {author} {\bibinfo {author} {\bibfnamefont {Y.}~\bibnamefont
  {Sun}}, \bibinfo {author} {\bibfnamefont {Z.}~\bibnamefont {Xia}}, \ and\
  \bibinfo {author} {\bibfnamefont {U.~S.}\ \bibnamefont {Kamilov}},\
  }\href@noop {} {\bibfield  {journal} {\bibinfo  {journal} {Archive}\ ,\
  \bibinfo {pages} {arXiv:1803.06594}} (\bibinfo {year} {2018})}\BibitemShut
  {NoStop}%
\end{thebibliography}%


\clearpage

\widetext

\begin{center}
\Large
\textbf{Supplementary material}
\normalsize
\end{center}
\vspace{-0.7cm}

\section{SLM calibration}

\begin{figure}[hb!]
\includegraphics{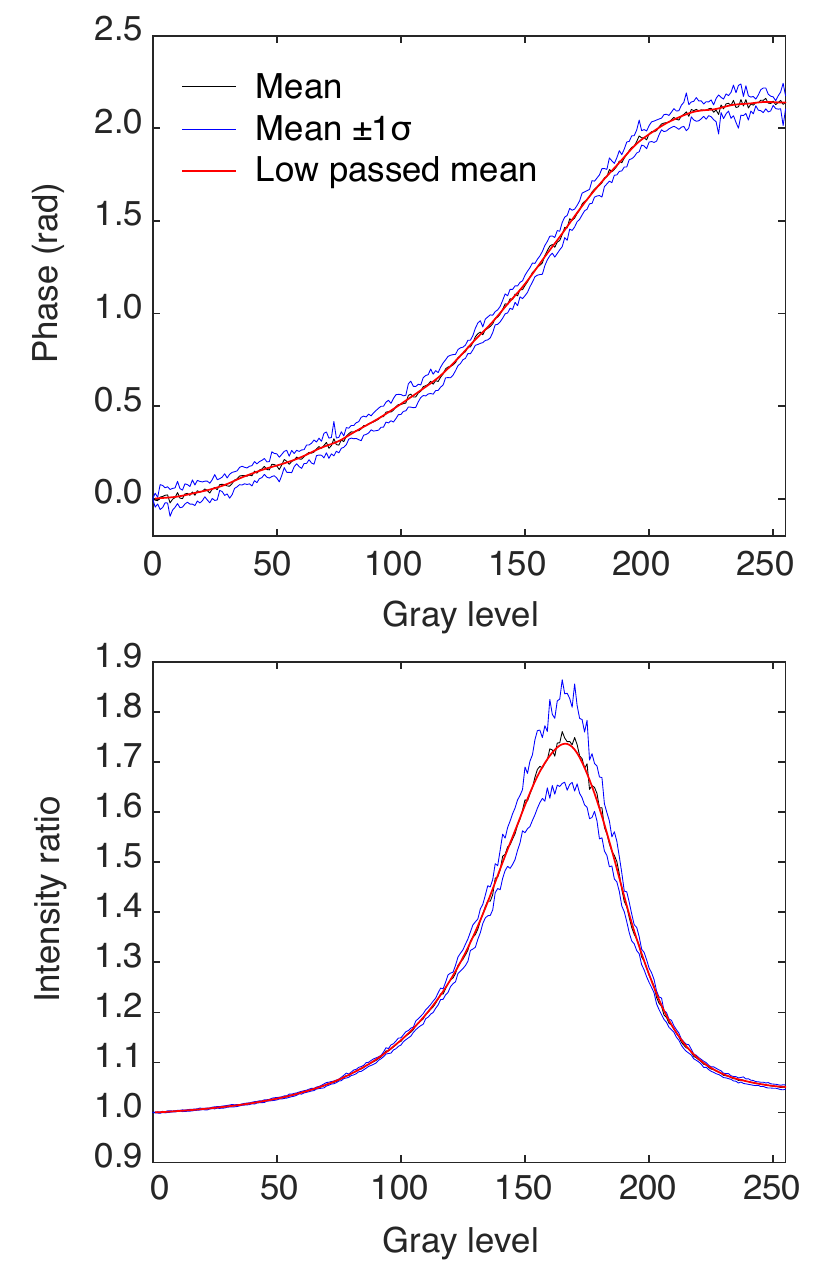}
\caption{\label{fig:slm} SLM calibration curves. Phase shift (a) and intensity ratio (b) as a function of gray level. On each graph, the central solid thin black line is the mean over 10 measurements, the thinner outer blue lines indicate $\pm$ 1 standard deviation from the mean. The thick red lines are the low passed version of the mean and were used as final calibration curves.}
\end{figure}

The complex transmittance of the SLM was measured for the particular position of polarizers P1 and P2 (see Fig.~\ref{fig:opt_setup} in the main text) by using a Mach-Zehnder interferometer and performing off-axis holographic measurements. The camera was placed in the image plane ($\Delta z = 0$) and a beam splitter was placed after lens L2 to obtain a reference beam, which was then recombined with the signal using a second beam splitter between lens L4 and the detector. The SLM is addressable as an external monitor and takes 8 bits integer values (called gray levels hereafter). The SLM was equally divided into two parts with a constant 0 gray level was assigned to the first part serving as a reference phase. The second part was assigned another constant gray level and the phase difference and amplitude ratio between the two parts were measured for each value of the gray level. The results are shown in Fig.~\ref{fig:slm}. The ground truths serving as the base for comparison in the correlation plots of Fig.~\ref{fig:correlations} were computed using the thick red curves of Fig.~\ref{fig:slm}.

\section{Photon count calibration}

The average number of photons per detector pixel quoted in the main text and in Table~\ref{tab:noiselevels} is based on measurements performed with a Silicon photodetector (Newport). The attenuation factor of the filter set used for each experiment was measured at the laser wavelength and is summarized in Table~\ref{tab:photon_calibration}. The absolute power of the whole illumination beam was measured between filter F2 and lens L4 of Fig.~\ref{fig:opt_setup} for the filter set of experiment 1. The powers for the other experiments were calculated using the filter attenuations measured in the first step. L4 induces a loss of $1.5\%$. The Silicon detector was large enough to capture the whole beam. However, the camera sensor being smaller than the beam, only a fraction of the power was captured. This fraction was calculated by fitting the incident beam measured by the camera with a model of the incident beam. The following beam model for the beam intensity $I$ was used, in polar coordinates:
\begin{equation}
	I(r, \theta) = J_0^2\left(a_0\frac{r}{R}\right)
\end{equation}

\noindent where $J_0$ is the $0^{\text{th}}$ order Bessel function of the first kind, $a_0$ the argument of the Bessel function $J_0$ at which the first minimum is reached, and $R$ the radius of the incident beam in the image plane. From this model, we obtain a radius $R = 8.5$mm. Note that only the central lobe of the Airy pattern from filter F1 was used (and whose power was measured) as an illuminating beam, the outer rings were blocked with a hard aperture set to the radius of the first intensity minimum. The fraction of the incident beam power captured by the camera detector is 69\%. The fourth column of Table~\ref{tab:photon_calibration} contains the number of photoelectrons generated in the detector, {\it i.e.} after the incident photon number had been multiplied by the quantum efficiency of the detector. Because the photoelectrons are the effectively measured quantity the photoelectron count is referred to as the photon count in Table~\ref{tab:noiselevels} of the main text. As already mentioned inthe main text, because the illumination beam does not have a constant intensity, the actual time-averaged photon count varies spatially.

\begin{table}[h]
\caption{\label{tab:photon_calibration} Photon energy $E_0 = hc/\lambda = 3.139\times 10^{-19}$J. Integration time = 2ms. Quantum efficiency = 0.6. Total number of pixels = 1,006,008.}
\begin{tabular}{|c|c|c|c|} \hline
Experiment & Filter set attenuation factor & Total beam power (W) & Average photoelectron count per pixel \\ \hline\hline
1 & $1.00 \times 10^2$ & 4.0$\times 10^{-7}$ & 1050 \\ \hline
2 & 1.23$\times 10^3$ & 3.3$\times 10^{-8}$ & 85 \\ \hline
3 & 2.38$\times 10^3$ & 1.7$\times 10^{-8}$ & 44 \\ \hline
4 & 1.06$\times 10^4$ & 3.8$\times 10^{-9}$ & 9.9 \\ \hline
5 & $1.00 \times 10^5$ & 4.0$\times 10^{-10}$ & 1.1 \\ \hline
6 & 4.15$\times 10^5$ & 9.6$\times 10^{-11}$ & 0.25 \\ \hline
\end{tabular}
\end{table}

\section{Simulations}

\begin{figure}
\includegraphics{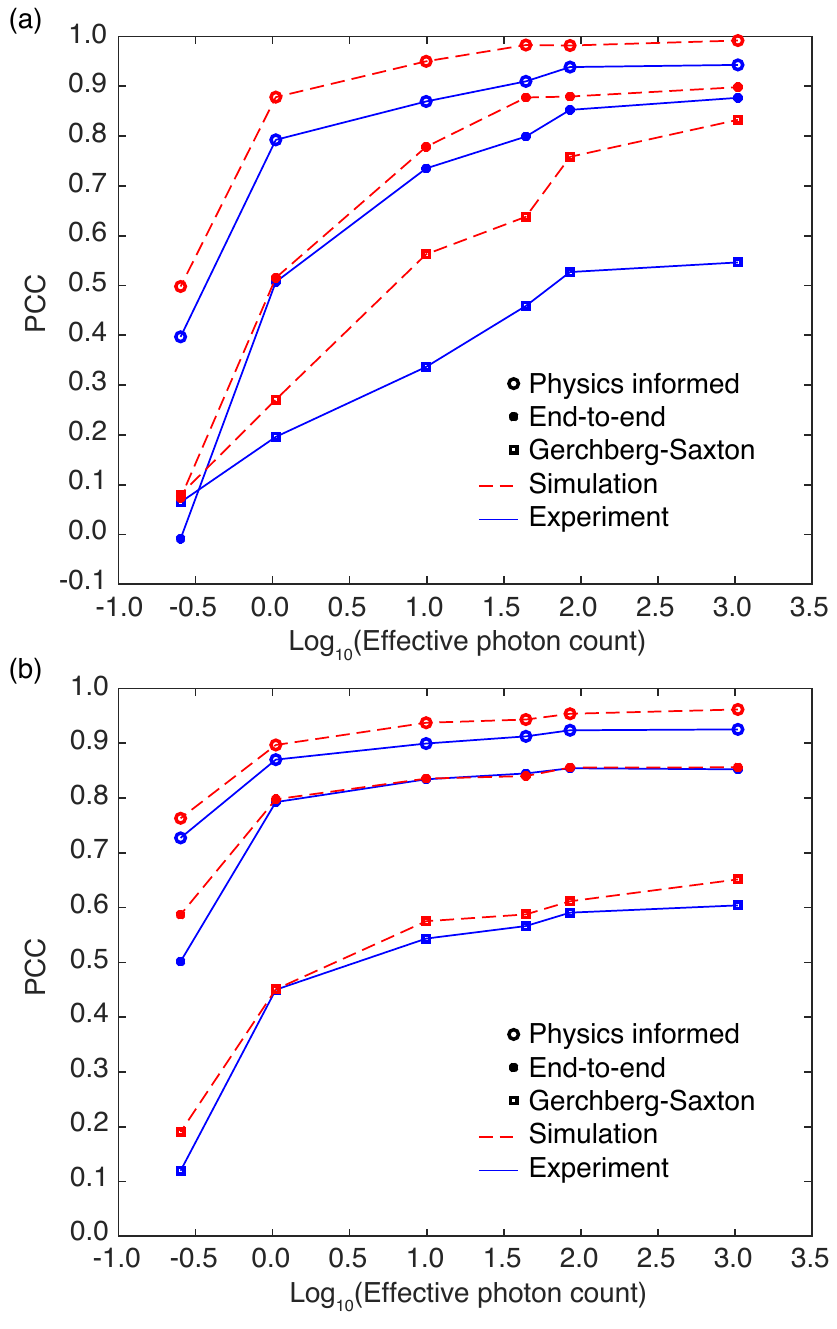}
\caption{\label{fig:simulations} Comparison of the correlation coefficients of reconstructions from experimental and simulated data. (a) IC layout data, (b) ImageNet data. The error bars are not shown to avoid overloading the graph.}
\end{figure}

The experimental acquisition process was simulated numerically. Synthetic camera measurements were generated with the same size and bit depth than the experimental measurements. The ground truths were calculated from the original 8 bit images by using the same SLM calibration curves (Fig.~\ref{fig:slm}) used for the experimental data. Fresnel propagation was computed using the fast Fourier transform. A noiseless diffraction pattern was computed in the detector plane and normalized so that the intensity corresponds to the average photon flux at each pixel $\phi$, for each nominal photon budget. For each pixel, the noisy signal $s$ from the detector was computed as follows:
\begin{eqnarray}
		p_c &=& \xi(\phi T) \\
		e_c &=& Q p_c \\
		s &=& s_0 + g G F e_c + \eta(G, T)
\end{eqnarray}
\noindent where $T$ is the integration time, $\xi(\lambda)$ is a random variable following the Poisson distribution with parameter $\lambda$, $p_c$ the photon count per pixel, $Q$ the quantum efficiency, $e_c$ the effective photon count or the generated electron count, $g$ is the camera pre-amplifier gain, $G$ is the EM gain, $F$ is a normally distributed random variable accounting for the noise in $G$ (excess noise factor) and $\eta$ is a normally distributed random variable accounting for the measured dark noise for the corresponding EM gain and exposure time. The correlations coefficients of the reconstructions with the ground truths are shown in Fig.~\ref{fig:simulations} together with the experimental data for comparison.

\end{document}